\documentclass[12pt,letterpaper,notoc]{JHEP3}
\newcommand{\beq}{\begin{equation}}                            
\newcommand{\eeq}{\end{equation}}
\def\bqa{\begin{eqnarray}}
\def\eqa{\end{eqnarray}}   
\preprint{UPR-0973-T}
\title{Implications of Gauge Unification for Time 
Variation of the Fine
Structure Constant}
\author{Paul Langacker\footnote{On leave at the
School of Natural Sciences, Institute for Advanced Study, Einstein Drive, Princeton, NJ 08540}, 
Gino Segr\`{e} and Matthew J. Strassler\\ \
\\
{\small{\it Department of Physics and Astronomy,
University of Pennsylvania}}\\
{\small{\it Philadelphia, PA 19104}}}

\abstract{
Unification of the gauge couplings would imply that time variations of the fine
structure constant are accompanied by calculable and very significant time
variations in the QCD scale parameter $\Lambda _{QCD}$. Since $\Lambda _{QCD}$
is the dominant factor in setting the
hadron masses, estimates made by simple variations
of the fine structure constant may not provide meaningful limits.
There may also be related variations in Yukawa couplings and the electroweak
scale. Implications for the 21 cm hyperfine transition,
big bang nucleosynthesis,
and the triple alpha process are discussed.  We find that the first
of these already provides strong constraints on the underlying theory.
It is emphasized more generally
that time (and space) variations of fundamental couplings
and their correlations
may be a significant probe of ultra-high-energy
physics.}

\begin{document}


Slow temporal variations of coupling constants are conceivable in many
theories, {\it e.g.}, those in which their values are related to
expectation values of scalar fields~\cite{wdp}.  (Naturally, spatial
variations may also occur~\cite{barrow}.)  This possibility has been
of considerable interest since first proposed by Dirac~\cite{Dirac}
over sixty years ago.  Webb {\it et al.}~\cite{webb} have recently
reported that that the fine structure constant $\alpha $ displays time
variation. The group reports a change $\frac{\Delta \alpha }{\alpha
}=-(0.72\pm 0.18)\times 10^{-5}$ over the redshift range $0.5 < z <
3.5,$ where $\Delta \alpha$ is $\alpha$ at large redshift minus
$\alpha$ at the present day.  In this note we will discuss how this
result, and other constraints, fit together if the coupling constants
of the strong, weak and electromagnetic interactions are unified.

 From a purely observational point of view, all fundamental constants,
from the cosmological constant on down, should be viewed as
potentially having spatial and time variation of an unknown sort.
More interesting than the particular variation of any individual couplings
would be the correlations in the variations of {\it different}
couplings.  Such correlations potentially will tell us how the
variations are being induced, and therefore give insight into the
underlying theory.  In short, temporal and spatial variations in
coupling constants may be a little-recognized probe of
ultra-high-energy physics.

Webb {\it et al.}'s \cite{webb} data is based on the absorption
spectra of distant quasars, but several constraints exist as well at
other redshift values~\cite{Chiba}. The best laboratory limit is
$\left| \frac{\Delta \alpha }{\alpha }\right| < 1.4\times 10^{-14}$
over a period of 140 days~\cite{prestage}.  The best geological limit
comes from the naturally occurring nuclear reactor at Oklo in Gabon;
recent re-analyses of this data~\cite{oklo}, under the assumption that
only $\alpha$ varies, and that $\frac{\dot{\alpha}}{\alpha}$ is constant,
lead to a limit of $\frac{\dot{\alpha}}{\alpha} =
(-0.2 \pm 0.8) \times 10^{-17}/{\rm yr}$ over two billion years.

There are other ongoing or planned observations which will provide
additional limits.  Fluctuations in the cosmic microwave background
resulting from changing the ionization history of the universe could
lead~\cite{cmb} to a measurement with experimental sensitivity of
$\left| \frac{\Delta \alpha }{\alpha }\right| < 10^{-2}-10^{-3}$ at
$z\sim 1000.$ Nucleosynthesis~\cite{bbn} places bounds on the
variations of $\alpha $ with roughly the same order of magnitude,
though at a much larger redshift, $z\sim 10^{9}-10^{10}$, again
assuming that the only variation is in $\alpha$.

If one assumes that $\frac{\dot{\alpha}}{\alpha}$ remains
approximately constant throughout the history of the universe, the
resultant value from the Webb {\it et al.}~\cite{webb} data is 
$\frac{\dot{\alpha}}{\alpha}
\sim 10^{-15}/{\rm yr},$ which is two orders of magnitude above the
Oklo bound although compatible with the other limits.  There is,
however, no particular reason to assume constant time variation, so
until the data suggests that $\frac{\dot{\alpha}}{\alpha}$ 
is constant, there is no clear
contradiction between Oklo and the new observations.

The time-variation of other quantities can 
also be measured using absorption lines.  Cowie
and Songaila~\cite{Cowie} constrain $X \equiv \alpha^2 g_p m_e/M_p$,
where the proton magnetic moment is $e g_p/2M_p$, from the 21 cm
hyperfine line in hydrogen at $z \sim 1.8$.  Similarly, Potekhin {\it
et al.}~\cite{Potekhin} limit $Y \equiv M_p/m_e$ from molecular
hydrogen clouds at $z = 2.81$:
\beq
\frac{\Delta X}{X} = (0.7 \pm 1.1) \times 10^{-5}, \ \ \ \ 
\frac{\Delta Y}{Y} = (8.3^{+6.6}_{-5.0}) \times 10^{-5}.
\label{XYexp}
\eeq

The main point of this paper\footnote{Some of the issues in this paper
 have been considered
in somewhat similar contexts, for example, in \cite{previous}}
 is to emphasize that the various
fundamental constants are likely to vary simultaneously and in a
correlated way that depends on the underlying physics, and that
studies of such systems as the Oklo reactor, nucleosynthesis, or
molecular/atomic absorption lines should take this into account.  We
will illustrate this by one especially well-motivated example, i.e.,
by theories in which the gauge couplings unify at some large scale
(consistent with the data~\cite{polonsky}).  Then a change in $\alpha$
will be accompanied by a much larger change in the strong interaction
coupling and mass scale.  For definiteness, we consider a grand
unified gauge theory such as $SU(5)$ with a single coupling constant
$\alpha _{G}$.  Similar considerations would apply to many string
theories which compactify directly to the standard
model.\footnote{There is an order of magnitude discrepancy between the
observed and expected unification scales in the simplest such
theories, but that is only a 10\% effect in $\log M_G/M_Z$.  This
could be accounted for by threshold or other effects, and would not
much affect our conclusions.}  Unified theories involving more
complicated breaking patterns, exotic matter, higher Ka\v c-Moody
levels, or large extra dimensions would differ in the details but not
in the qualitative features.

Unification presumably occurs at some large scale $M_{G},$ typically of
order $3\times 10^{16}$ GeV. At lower scales the group breaks down to $SU(3)
\times SU(2) \times U(1)$ characterized by couplings $\alpha _{i}$,
$i=1,2,3,$  which however tend to $\alpha _{G}$ as the scale at
which they are measured nears $M_{G}.$ 
For instance, in the MSSM (Minimal
Supersymmetric Standard Model), the couplings run so that at the
electroweak scale $M_{Z}$ they
are given by
\beq
\alpha _{i}^{-1}(M_{Z})=\alpha _{G}^{-1}+b_{i}t_{G}
\label{rgeeqn}
\eeq
where $t_{G}=\frac{1}{2\pi }$ $\ln \frac{M_{G}}{M_{Z}}\approx 5.32$ and $\alpha
_{G}^{-1}\approx 23.3.$ The $b_{i}$ are fixed by the relative particle
multiplets; in the MSSM one has $b_{i}=\left( \frac{33}{5},1,-3\right).$
Between $M_{G}$ and $M_{Z}$ the
$\alpha _{i}$ rather than $\alpha$ are the relevant couplings.
Below $M_{Z}$ the standard model group $SU(3)
\times SU(2) \times U(1)$ is further broken to $SU(3)\times U(1)_{ELM}$,
where $U(1)_{ELM}$ is the electromagnetic gauge group.
This leads  to a value of the electromagnetic coupling at scale $M_{Z}$
of
\beq
\alpha ^{-1}\left( M_{Z}\right) =\frac{5}{3}\alpha _{1}^{-1}\left(
M_{Z}\right) +\alpha _{2}^{-1}\left( M_{Z}\right)
=127.9.
\eeq

Now imagine that $\alpha _{G}^{-1}$
is the vacuum expectation value of some field $\phi (t)$ which
is a slowly varying function of time. (For example,
$\phi (t)$ might parametrize the
change in the volume of some compact extra dimensions.)
This in turn will induce correlated variation, at any given
energy scale below $M_G$, in the $SU(3)
\times SU(2)\times U(1)$ coupling constants. 
To find the relationship between the couplings,
we make a second assumption (which we will relax later)
that the changes in $\alpha _{i}^{-1}$ are dominated by the 
changes in $\alpha _{G}^{-1}$ --- in
other words, that
 corrections to $t_G$ and to threshold effects, etc., are relatively
unimportant.\footnote{Clearly we are also assuming 
that the back-reaction of a
varying $\alpha_G$ on the cosmological constant is not large; we will comment
on this below.}  We also neglect higher order corrections and
assume (correctly, in the majority of models) that 
$\Delta \alpha (M_Z)/\alpha (M_Z) \simeq \Delta \alpha/\alpha$,
where $\alpha\equiv\alpha (0)\sim 1/137$ is the conventional low-energy
fine structure constant.  We then have
\beq
\frac{\Delta \alpha _{i}}{\alpha _{i}}=-\frac{\Delta \alpha _{i}^{-1}}{
\alpha _{i}^{-1}}=-\frac{\Delta \alpha _{G}^{-1}}{\alpha _{i}^{-1}},
\eeq
so that at scale $M_{Z}$
\beq
\frac{\Delta \alpha }{\alpha }=
- \frac{8}{3}\frac{\Delta \alpha _{G}^{-1}}{\alpha ^{-1}}
=\frac{8}{3}\frac{
\alpha }{\alpha _{G}}\frac{\Delta \alpha _{G}}{\alpha _{G}}
\sim 0.49\frac{\Delta 
\alpha _{G}}{\alpha _{G}}.
\eeq
Similarly, the $SU(3)$ group's coupling, usually called $\alpha _{S}$
instead of $\alpha _{3},$ satisfies, again at scale $M_{Z}$,
\beq\label{alphasMZ}
\frac{\Delta \alpha _{S}}{\alpha _{S}}\simeq \frac{3
}{8}\frac{\alpha _{S}}{\alpha }\frac{\Delta \alpha }{\alpha }\sim 5.8\frac{
\Delta \alpha }{\alpha }
\eeq
{\it i.e.}, 
the proportional variation in the strong subgroup's coupling is almost
six times as great as that of $\alpha$, when measured at the scale $M_{Z}$, 
where $\frac{\alpha _{S}}{\alpha }\simeq 15.4$.

The physical quantities we are most directly interested in are the
associated relative changes in the neutron and proton masses and, to
the extent this is possible to estimate, the changes in nuclear
binding energies. The nucleon masses are largely fixed by the QCD
scale $\Lambda _{QCD}$. The expression we are most interested in is
therefore the relation of $\frac{\Delta \Lambda _{QCD} }{\Lambda
_{QCD}}$ to $\frac{\Delta \alpha }{\alpha }$ (up and down quark masses
make only a minute contribution to nucleon masses).  $\Lambda _{QCD}$
is approximately the scale at which $\alpha _{S}$ diverges, $\alpha
_{S}^{-1}\left( \Lambda _{QCD}\right) =0.$ To see how changes in this
scale are related to changes in $\alpha ,$ we must study the further
running of $\alpha _{S}$ from the scale $M_{Z}$ down to $\mu .$ This
change is again given by the renormalization group equations 
\beq
\alpha _{S}^{-1}(\mu )=\alpha _{S}^{-1}(M_{Z})- \frac{b_{3}^{\rm
SM}}{2\pi }\ln \left( \frac{\mu }{M_{Z}}\right) 
\eeq 
where $\alpha_{S}^{-1}(M_{Z})=8.33$ and $b_{3}^{\rm SM}=-11+\frac{2}{3}n_{F}$,
where $n_{F}$ is the number of quark flavors light compared to $\mu$.

Using the above equation, the value
of $\mu $ at which $\alpha _{S}^{-1}(\mu )\rightarrow 0$ is given by
\beq
\Lambda _{QCD}^9\sim M_{Z}^{23/3}m_b^{2/3}m_c^{2/3}\exp \left(-\frac{
2\pi}{ \alpha _{S}\left( M_{Z}\right)}\right).
\eeq
where $m_b, m_c$ are the bottom and charm quark masses, whose
subleading dependence on $\alpha_S$ we are ignoring.
This leads immediately to the desired relation
\beq
\frac{\Delta \Lambda _{QCD}}{\Lambda _{QCD}}
=-\frac{2\pi }{9}\Delta 
\alpha _{S}^{-1}\left( M_{Z}\right)
=\frac{2\pi }{9\alpha _{S}}\frac{
\Delta \alpha _{S}}{\alpha _{S}}
\eeq
where $\alpha _{S}$ is measured at $M_{Z.}$ Using (\ref{alphasMZ}),
 we finally arrive at a prediction 
\beq\label{DeltaQCD}
\frac{\Delta \Lambda _{QCD}}{\Lambda _{QCD}}\sim 34\frac{\Delta \alpha
\left( M_{Z}\right) }{\alpha \left( M_{Z}\right) }
\sim 34\frac{\Delta \alpha}{\alpha  }.
\eeq
(The coefficient has a theoretical uncertainty at the level of perhaps 
twenty percent.)
This suggests that, within the framework of a unified
theory and assuming the nucleon mass scale is set by $\Lambda _{QCD},$
a variation such as the one suggested by Webb {\it et al.}~\cite{webb} of
$\frac{\Delta \alpha }{\alpha } =-0.72\times 10^{-5}$ should be
accompanied by a shift in $\frac{\Delta M_{p} }{M_{p}}\sim -25\times
10^{-5},$ clearly a very significant correction.

Before discussing the implications, let us consider possible variations
in other physical constants. These are likely to occur along with
those in the gauge couplings, although the specific relation cannot
be obtained without a more specific theory. However, it is useful
to parametrize them and consider likely possibilities.

Only dimensionless quantities such as coupling constants or
ratios of masses are physically significant. We take the point of
view that the unification scale $M_G$ is simply a
reference scale relative to which other masses such as $\Lambda_{QCD}$
are expressed.  Indeed, the formula (\ref{DeltaQCD}) should
more properly be interpreted, in this language, as a statement
about the variation in the quantity $\Lambda_{QCD}/M_G$.  Later
we will have to consider the variation in the
Newton constant $G_N \propto M_{pl}^{-2}$ relative to $M_G$.

We next consider the (running) Yukawa coupling $h_a$ of fermion
$a$, which is related to its mass
by $m_a = h_a v$, where $v \sim 246$ GeV is the electroweak
scale. (For the quarks, the physical mass is $\sim (m_q + M_{QCD})$,
where  $M_{QCD} \sim \Lambda_{QCD}$, the latter contribution dominating
for $u$ and $d$.)
We expect that since the gauge coupling is varying at the unification scale,
the same will be true, to some degree, for the Yukawa couplings.  Since
the size of this effect is model-dependent, we parametrize our ignorance
by introducing unknown constants $\lambda_a$, where
\beq
\frac{\Delta h_{a}(M_G)}{h_{a}(M_G)} \equiv \frac{\lambda_a}{2}
\frac{\Delta \alpha_G}{\alpha_G} \sim \lambda_a 
\frac{\Delta \alpha}{\alpha}.
\eeq

The running of $h_a$ is calculable in a given model. For the
light quarks and leptons, the running is dominated by the gauge
contributions, yielding 
\beq
\frac{h_{a}\left( M_{G}\right) }{h_{a}\left( M_{Z}\right) }%
=\prod_{i=1,2,3}\left[ \frac{\alpha _{G}\left( M_{G}\right) }{\alpha
_{i}\left( M_{Z}\right) }\right] ^{\frac{b_{a;i}}{2b_{i}}}
\eeq
where in the MSSM the anomalous dimensions are given by
$b_{u;i}=\left( \frac{-13}{15},-3,\frac{-16}{3}\right) ,b_{d,i}=\left(
\frac{-7}{15},-3,\frac{-16}{3}\right) $ and $b_{e;i}=\left( \frac{-9}{5}%
,-3,0\right).$
This implies
\beq
\frac{\Delta h_a(M_Z)}{h_a(M_Z)} \sim \left(\lambda_a - \sum_i 
  b_{a;i} \alpha_i t_G\right)
\frac{\Delta \alpha}{\alpha},
\eeq
where $\alpha_i$ is evaluated at $M_Z$. One finds
$ - \sum_i b_{a;i} \alpha_i t_G \sim 4.0$ 
for $a=u$ or $d$, and $0.70$ for $a=e$.
That is, there is a moderate  magnification of the quark Yukawas
from the gauge corrections (mainly from $\alpha_S$), but a much
smaller effect for the electron.  One might expect that
this effect for the quarks gets further magnified as one runs from $M_Z$ 
down to the quark mass.  This is true for the bottom and
charm Yukawa couplings.  However, for the light quarks, the running should
be taken down to a scale of order $\Lambda_{QCD}$.  When this
is done, and the variation of $\Lambda_{QCD}$ is accounted for,
it turns out there is a cancellation and the effect is actually
smaller.  Given the inherent theoretical uncertainties in our computations,
we have chosen to ignore this small shift, and will
simply use the $\lambda_a$ for the light quarks.

More important are the changes in the electroweak scale $v$, which are
tied to the scale of supersymmetry breaking in most supersymmetric
models. For example, in the traditional supergravity mediated models
there are various soft supersymmetry breaking masses (scalar masses,
sfermion masses, and other bilinear and cubic scalar terms), which are
usually assumed to have the same order of magnitude $m_{soft}$ at the
Planck scale.\footnote{We are assuming that $\mu$, the supersymmetric
Higgs mass, is comparable to the soft parameters, whether it is
elementary or dynamically generated.}  Although these soft parameters
run (and may change sign for scalar mass-squares), they are generally
of the same order of magnitude at the weak scale.  Typically, one
finds that $v^2$ (and the inverse $G_F^{-1}$ of the Fermi constant)
scales as $m_{soft}^2/\alpha_{weak}$, where $\alpha_{weak} =
\frac{3}{5} \alpha_1 + \alpha_2$, and that the corresponding $W$ and
$Z$ masses scale as $m_{soft}$. The underlying mechanism for breaking
supersymmetry in the hidden sector and therefore generating $m_{soft}$
is unknown. We shall parametrize our
ignorance by introducing \beq \frac{\Delta v}{v} \equiv \kappa
\frac{\Delta \alpha}{\alpha}.
\label{kappa}
\eeq
Then,
\beq
\frac{\Delta m_a}{m_a} \sim  (\lambda_a
 - \sum_i b_{a;i} \alpha_i t_G  + \kappa)
 \frac{\Delta \alpha}{\alpha} .
 \label{fermionmass}
 \eeq

In many
supersymmetry-breaking models, a dynamical strong-coupling scale sets
an intermediate supersymmetry breaking scale $M_I$, which in
turn feeds into $m_{soft}$.  Typically this relation
takes a form $m_{soft} \sim M_I^n/M_G^{n-1}$, $n$ positive. It is
easy to see that if the supersymmetry-breaking dynamics
is also unified with the standard model, with coupling
constant $\alpha_G$ at $M_G$, then for any $n$
\beq
\frac{\Delta m_{soft}}{m_{soft}}
\sim
- \frac{\Delta  \alpha_G}{\alpha_G} \ln \frac{m_{soft}}{M_G}
  \sim
  35 \frac{\Delta  \alpha_G}{\alpha_G}
  \sim
  70 \frac{\Delta \alpha}{\alpha}
\eeq
or, in short, $\kappa \sim 70$.  (Note that there is no corresponding
reason to expect the $\lambda_a$ to be large.) 
However, the coupling constant of the
supersymmetry-breaking physics may be set in other ways, in which case
$\kappa$ is completely unconstrained.  

One can now go back and investigate the correctness of our original
assumption of considering only the change in $\alpha_G^{-1}$ and not
that in $t_G$ in (\ref{rgeeqn}).  It is easily shown that the effect
of the change in $t_G$ induced by $\Delta v$ is not small when
$|\kappa|$ is of order 10 or larger.  For example, the second
and third expressions in   (\ref{alphasMZ}) acquire correction
factors $(1-10\frac{\alpha}{\pi}\kappa)\sim (1-0.025\kappa)$.
It is appropriate, then, to recalculate the variation of $\Lambda_{QCD}$
to see whether we have left out a large effect.  We find
\beq\label{DeltaQCDb}
\frac{\Delta \Lambda _{QCD}}{\Lambda _{QCD}}\sim 
34(1+.005\kappa)\frac{\Delta \alpha}{\alpha  }.
\eeq
where the coefficient of $\kappa$ is uncertain at the level of twenty
percent due to ambiguities in the one-loop renormalization prescription.
Note that the effect of $\kappa\sim 70$ is substantial but
not overwhelming.  The uncertainties in its effect are comparable
to or below the errors in \cite{webb}, as we will see.

We finally mention that changes in the gauge couplings may also be
correlated with changes in the cosmological constant $\Lambda$.  Indeed,
it has recently been argued that fine-tuned cancellations involving
radiative contributions would be upset by a time variation in
$\alpha$, with enormous effect~\cite{Banks}. Since the smallness of
$\Lambda$ is the outstanding puzzle in particle physics, it is
difficult to know how it is affected.  Naively, quantum field theory
and classical gravity would suggest the back-reaction due to varying
couplings would be enormous, but we already know that effective field
theory reasoning is wrong, and do not know why.  Since there is
neither theoretical nor experimental guidance on this point, we choose
to assume the cosmological constant is not affected; whether this is
correct is, in our view, still an experimental question.

As a first application, let us consider the quantities
$X \equiv \alpha^2 g_p m_e/M_p$ and $Y \equiv M_p/m_e$.
For the nucleon masses, one has
\bqa
M_p & = & M_{nuc} + 2 m_u + m_d + \alpha M_{elm} \nonumber \\
M_n & = & M_{nuc} +  m_u + 2 m_d,
\label{nucleons}
\eqa
where $ M_{nuc} \sim 3 M_{QCD}$  scales like $\Lambda_{QCD}$.
$\alpha M_{elm}$ is the electromagnetic
contribution contribution to $M_p$, with $M_{elm}$ also scaling
like $\Lambda_{QCD}$. The variations of $m_{u,d}$ and $m_e$ are
given in (\ref{fermionmass}). $m_{u,d}$ are typically
estimated to be a few MeV, so that $M_{p,n}$ are dominated by $ M_{nuc}$.
Hence, we expect
\beq
\frac{\Delta M_p}{M_p} \sim \frac{\Delta \Lambda_{QCD}}{\Lambda_{QCD}}.
\eeq
(This would have to be modified if $\lambda_a$ or $\kappa$ were enormous.)
The proton and neutron $g$ factors $g_{p,n}$ are well described in
the constituent quark model, where they are Clebsch-Gordan coefficients,
so we will ignore possible variation in $g_p$. Assuming
that $\lambda_u\approx\lambda_d\approx\lambda_e$, and denoting their
average value as $\lambda$, one then has
\bqa
\frac{\Delta X}{X} & \sim & \left(-32 + \lambda + 0.8\kappa \right)
  \frac{\Delta \alpha}{\alpha} \sim (23 \pm 6) \times 10^{-5}  \nonumber \\
\frac{\Delta Y}{Y} & \sim & \left( 34 - \lambda - 0.8\kappa \right)
  \frac{\Delta \alpha}{\alpha} \sim (-24 \pm 6) \times 10^{-5},
  \label{XY}
  \eqa
where the numerical values are obtained using the Webb {\it et al.}
value $\frac{\Delta \alpha}{\alpha} = (-0.72 \pm 0.18) \times 10^{-5}$
and $\lambda = \kappa = 0$.  (The theoretical errors in this
computation, which are slightly smaller than the experimental errors
but could straightforwardly be reduced by more careful calculation,
are not shown.)  One sees the large magnification implied by coupling
constant unification.  The values in (\ref{XY}) are to be compared
with the observational limits~\cite{Cowie,Potekhin}, given in
eq.~(\ref{XYexp}),
obtained at redshifts within the Webb {\it et al.} range.  Consistency of
(\ref{XY}) and (\ref{XYexp}) would require a rather delicate
cancellation, with $\lambda + 0.8 \kappa \sim 32$.  

We should emphasize that one ought not to conclude from this that the
result of \cite{webb} is inconsistent with unification of coupling
constants. Even if $\lambda+0.8 \kappa$ lies outside the range
$25-35$, one can conclude only that unification requires that the
variation observed by Webb {\it et al.}~is caused by physics at
distance scales long compared with $M_G^{-1}$, and preferentially
affecting electromagnetic or electroweak physics.  Our aim is to use
\cite{webb} to constrain the theory, not the reverse.

Let us now turn to another process which also could be strongly 
affected by coupling
variations: nucleosynthesis.  Consider the neutron-proton ratio in the early
universe, fixed approximately at the temperature $T_{F}$ at which the the
weak-interaction processes that interconvert neutrons and 
protons freeze out by
comparison to the expansion rate. The relative abundance is
\beq
\frac{N_{n}}{N_{p}}\simeq e^{\frac{-(M_n-M_p)}{T_{F}}},
\eeq
where by (\ref{nucleons})
the neutron-proton mass difference
$M_{n}-M_{p} \sim 1.29$ MeV  is given by
\beq
M_{n}-M_{p}=m_{d}-m_{u}-  \alpha M_{elm}.
\eeq
For definiteness, we will use the estimates~\cite{Gasser}
$m_{d}-m_{u} = 2.05 \pm 0.30$ MeV, and $\alpha M_{elm} = 0.76 \pm 0.30$
MeV.  The freeze-out temperature $T_{F} \sim 0.72$ MeV
is set roughly by equating the weak interaction rate at temperature $T$
to the universe's expansion rate. Expressing the Fermi constant $G_{F}$
in terms of the Higgs vacuum expectation value $v$ gives
\beq
\frac{T^{5}}{v^{4}}\simeq K\frac{T^{2}}{M_{Pl}}\Longrightarrow
T_{F}\simeq \left( K\frac{v^{4}}{M_{Pl}}\right) ^{\frac{1}{3}},
\eeq
where $K$ is a constant. Using (\ref{kappa}) and defining
\beq
\frac{\Delta M_{pl}}{M_{pl}}=\rho \frac{\Delta \alpha}{\alpha}  
\eeq 
(recall we are holding $M_G$ fixed,) we have
\beq
\frac{\Delta T_F}{T_F} = \left(\frac{4}{3} 
\kappa-\frac{1}{3}\rho\right) \frac{\Delta \alpha}{\alpha} .
\label{TF}
\eeq

The variation in $r \equiv N_n/N_p$ is given by
\beq
\frac{\Delta r}{r} = - \Delta \left( \frac{M_{n}-M_{p}}{T_{F}} \right).
\eeq
In addition to the explicit $\alpha $ dependence, there may be a much
larger variation due to $ M_{elm} $, and possibly the quark mass difference
and $T_F$. From (\ref{fermionmass}), (\ref{nucleons}), and (\ref{TF}) we find
\beq
\frac{\Delta r}{r} \sim 37 (1 - 0.08 \lambda - 0.01 \kappa -0.02\rho)
\left( \frac{\Delta \alpha}{\alpha} \right)_{\rm BBN}, 
\eeq
where the subscript BBN indicates that the quantity is to
be evaluated at the time of nucleosynthesis. Possible changes
in $r$ can be estimated from the $^4He$ abundance~\cite{Steigman},
with typical estimates implying
\beq
-0.1 < \frac{\Delta r}{r} < 0.02,
\eeq
implying
\beq
-0.003 < (1 - 0.08 \lambda - 0.01 \kappa -0.02\rho)
\left( \frac{\Delta \alpha}{\alpha} \right)_{\rm BBN} < 0.0005.
\label{bbnbound}
\eeq 
(Note that the coefficient could be negative for $\lambda + 0.8\kappa
\sim 32$.) If we assume a constant $\dot{\alpha}/\alpha$, this
implies the range $(-3 \times 10^{-13}/{\rm yr}, 5 \times
10^{-14}/{\rm yr})$ for $ (1 - 0.08 \lambda - 0.01 \kappa -0.02\rho)
\frac{\dot{\alpha}}{\alpha}$, 
which is consistent with the result of \cite{webb}. However,
other time dependences (e.g., a linear dependence on $1+z$) could give
very different results; and we already know that, for 
constant $\frac{\dot{\alpha}}{\alpha}$,
Oklo and the results of \cite{webb} are inconsistent (but see below.)
Thus it is vital to measure the time dependence of $\alpha$ at
moderate $z$ with high precision.

Leaving aside time variation for the moment, we may recall that the
actual values of fundamental couplings have themselves been the
subject of continued interest, particularly in light of the fact that
a universe like ours may only be possible for a narrow range of
parameters. This subject, commonly known by the catchword of the
``anthropic principle''~\cite{Hogan},
bears on the issue of varying couplings.  If the time
dependence of the couplings was such as to bring them outside the
range of allowed values for certain reactions, the anthropic principle
applied at an earlier era might forbid an otherwise acceptable time
variation.  

In particular, there are contexts in nuclear physics where
high sensitivity to varying coupling constants has been noted.  One
might wonder, then, whether the observed variation might in 
turn affect the observed properties of
stars, or affect the natural reactor at Oklo sufficiently to warrant a
reanalysis of results obtained from it.

Agrawal {\it et al.}~\cite{Agrawal} have discussed at length, in a variety of
settings, the anthropic principle's constraints on the value of $v$,
the Higgs vacuum expectation value. Recent extensions of this analysis
have focused on the much tighter constraints placed by the existence
of the triple-alpha process $^{4}He+^{8}Be\longrightarrow ^{12}C^{*}$
where $^{12}C^{*}$ is an excited state of carbon, 7.6 MeV above the
ground state. The existence of this resonance and the non-existence of
a resonance below threshold in oxygen ensures that (1) carbon is
produced in stellar interiors and (2) the carbon is not immediately
converted into oxygen. In other words the existence of carbon in our
universe is sensitively dependent on the location of nuclear
resonances in both carbon and in oxygen. The subject becomes
particularly interesting for our considerations because Livio et
al.~\cite{Livio} and more recently Oberhummer {\it et al.}~\cite{oberhummer}
have shown how sensitively the energies of these resonances depends on
the basic structure of the nucleon-nucleon potential.  As an
illustration, consider carbon production in stars, which has been
occurring in stars for many billions of years: the time variation of
$\alpha $ cannot be such as to imply a change in the nucleon-nucleon
potential by a quantity sufficient to shut down carbon production.
There is some question of how much of a change in the nucleon-nucleon
potential is acceptable, but it is apparently of order 1\%.

A shift in $\Lambda_{QCD}$, as in equation (\ref{DeltaQCD}), 
is by itself not very important, because
the main features of nuclei (nucleon masses, nuclear potential depths)
shift along with it. At leading order, $\Lambda_{QCD}$ is the only
scale affecting QCD physics, so that all dimensionless quantities are
shift-invariant.  However, this scaling is broken by the quark masses;
and in stars, it is also broken by any associated changes in the stellar
environment, such as the core temperature, which might be
induced by changes in the various parameters (including $M_{pl}$.)  
In particular, unlike most hadronic
masses, the pion mass does not scale like $\Lambda _{QCD}$.  This is
because of its special role as a pseudo-Goldstone boson of an
approximate global chiral $SU(2)$ symmetry of the strong interactions,
which is explicitly broken by the small quark mass terms. One has
$m_{\pi }^{2}\sim f_{\pi }\left( m_{u}+m_{d}\right) $ with the pion
decay constant $f_{\pi } \sim \Lambda _{QCD}$.  At some level this
effect will feed into other hadronic masses, such as that of the
$\rho$, and modify their linear scaling, and will also affect such
dimensionless quantities as the pion-nucleon coupling. However, the
dominant effect should be on the range of the pion exchange potential,
which effectively shifts the strength by a factor of $-\Delta
(m_\pi/M_{had})$, where $M_{had}$ is a typical hadronic scale. 
If we assume for illustration that this must be $< 1$\%, one obtains
\beq
\left| \frac{\Delta (m_\pi/M_{had})}{(m_\pi/M_{had}}\right|
\sim \frac{1}{2}
\left| \frac{\Delta (m_q/\Lambda_{QCD})}{m_q/\Lambda_{QCD}}\right|
\sim  \frac{1}{2}
\left| (-34+\lambda + 0.8\kappa) \frac{\Delta \alpha}{\alpha} \right|
< 0.01.
\eeq
The limit on $\frac{\Delta \alpha }{\alpha }$ is then of order
$10^{-2}-10^{-3}$ over a period of five billion years, a constraint
certainly compatible with the Webb {\it et al.}~\cite{webb}
observation.  (This simple estimate is consistent, when appropriately
translated, with the more thorough calculation of Jeltama and Sher
\cite{Jeltema}, who considered the changes induced in the potential by
variations in $v$, keeping $\alpha_S(M_Z)$ fixed.)  A full
calculation of the effect of varying $m_\pi/M_p$, using a realistic
nuclear potential model along the lines of \cite{Jeltema}, and taking
into account possible variations in the stellar environment, would be
useful.

Meanwhile, the Oklo reactor data have so far been analyzed in detail
only for the effects of a direct variation of $\alpha$.  Changes in
the effective nuclear potential, especially from the change in
$m_\pi/M_{had}$, could have a significant effect on our understanding
of its implications.  In particular, the statement that the combined
results of Webb {\it et al.} and those of Oklo are inconsistent with
constant $\frac{\dot{\alpha}}{\alpha}$ 
assumes that only $\alpha$ is varying; if the nuclear
physics is also varying, then no firm conclusion is possible at this
time.  A reanalysis of the Oklo reactor, aimed at obtaining more
accurately the constraints that it imposes on the space of coupling
constants, would therefore be welcome.  It is possible that this
reanalysis would put strong constraints on the variation of
$\Lambda_{QCD}/v$, as well as on $\alpha$, 
and these would be interesting to know.

To summarize,  we have considered the possibility that the variation of
coupling constants is caused by physics at very short distance scales,
where the gauge couplings may be unified.  If this is the case, one
should not treat the variation of the fine structure constant in
isolation.  One must simultaneously treat the variations in the strong
coupling; the running of coupling constants suggest these are very
significant. Within our framework a measured variation of an
expression of the form $\alpha \frac{m_{e}}{M_{p}}$, for example, is
due more to changes of $\frac{m_{e}}{M_{p}}$ than of $\alpha$.  We
have also calculated the changes induced in Yukawa couplings by
anomalous dimensions.  Parametrizing the changes in Yukawa couplings
at the unification scale by $\lambda$ and in the electroweak scale $v$
by $\kappa$, respectively, we find the Webb {\it et al.} observations are
consistent with other absorption line results (which depend on
$m_e/M_p$) only for specific and large values for $\lambda + 0.8\kappa$.
Large effects are also possible for the abundance of primordial
$^4He$, but these are at a much earlier period and cannot be directly
related to the absorption line results without a detailed model of the
time dependence.  A reanalysis of the Oklo reactor constraints and a
detailed study of the triple alpha process, taking into account these
types of effects, would be extremely useful.

More generally, the study of temporal and spatial variations in
fundamental constants is potentially a powerful probe of fundamental
physics. Both the time-variation of these couplings (which can be
constrained by combining, {\it e.g.}, nucleosynthesis, cosmic
microwave radiation, and absorption line data), and the correlations
amongst any observed variations at a fixed time, provide strong
diagnostics of the underlying theoretical structure.  It is important
that thorough and dedicated efforts to constrain such variation be
made.

As this work was completed, a paper by Calmet and Fritzsch
appeared~\cite{calmet} with considerable overlap.
Though there are differences
in the details, the two papers agree on the main point:
that coupling constant unification imples that a time variation in $\alpha$
be  accompanied by a much larger variation in strong
interaction parameters such as the nucleon mass, and that limits on
time variation need to be recalculated and re-interpreted.

All three authors wish to express their gratitude to the Department of
Energy for support through grant DOE-EY-76-02-3071.  One of us (PL)
was also supported by the W. M. Keck Foundation as a Keck
Distinguished Visiting Professor at the Institute for Advanced Study
and by the Monell Foundation.  PL thanks J. Bahcall,
F. Dyson and N. Seiberg,
and MJS thanks J. Erler, A. Naqvi and G. Shiu, for useful discussions.


\begin{thebibliography}{99}
\bibitem{wdp} See, for example,
J. D. Bekenstein, Phys. Rev. D {\bf 25}, 1527 (1982);
T.~Damour and A.~M.~Polyakov, Nucl.\ Phys.\ B {\bf 423}, 532 (1994);
B.~A.~Campbell and K.~A.~Olive,
Phys.\ Lett.\ B {\bf 345}, 429 (1995);
E.~Witten, hep-ph/0002297;
K.~A.~Olive and M.~Pospelov, hep-ph/0110377.

\bibitem{barrow} J.~D.~Barrow and C.~O'Toole, astro-ph/9904116.

\bibitem{Dirac}  P. A. M. Dirac, Nature {\bf 139}, 323 (1937).

\bibitem{webb}  J. K. Webb {\it et al.},
Phys. Rev. Lett. {\bf 87}, 091301 (2001).

\bibitem{Chiba} For a general review of constraints on the variation
of fundamental constants, see
T. Chiba, gr-qc/0110118.

\bibitem{prestage} J.~D.~Prestage, R.~L.~Tjoelker and L.~Maleki,
  Phys. Rev. Lett. {\bf 74}, 3511 (1995).

\bibitem{oklo} T.~Damour and F.~Dyson, Nucl.\ Phys.\ B {\bf 480}, 37 (1996);
Y.~Fujii {\it et al.},
Nucl.\ Phys.\ B {\bf 573}, 377 (2000).

\bibitem{cmb}
S.~Hannestad, Phys.\ Rev.\ D {\bf 60}, 023515 (1999);
M.~Kaplinghat, R.~J.~Scherrer and M.~S.~Turner,
Phys.\ Rev.\ D {\bf 60}, 023516 (1999).

\bibitem{bbn}  E.~W.~Kolb, M.~J.~Perry and T.~P.~Walker,
Phys.\ Rev.\ D {\bf 33}, 869 (1986);
L.~Bergstrom, S.~Iguri and H.~Rubinstein,
Phys.\ Rev.\ D {\bf 60}, 045005 (1999);
P.~P.~Avelino {\it et al.},
Phys.\ Rev.\ D {\bf 64}, 103505 (2001).

\bibitem{Cowie}  L. L. Cowie and A. Songaila, Astrophys. J. {\bf 453},
596 (1995).

\bibitem{Potekhin}
A.~Y.~Potekhin {\it et al.},
Astrophys.\ J.\  {\bf 505}, 523 (1998).

\bibitem{previous}
W.~J.~Marciano,
Phys.\ Rev.\ Lett.\  {\bf 52}, 489 (1984);
C.~T.~Hill and G.~G.~Ross,
Nucl.\ Phys.\ B {\bf 311}, 253 (1988);
C.~T.~Hill, P.~J.~Steinhardt and M.~S.~Turner,
Phys.\ Lett.\ B {\bf 252}, 343 (1990).

\bibitem{polonsky} See, for example,
P.~Langacker and N.~Polonsky,
Phys.\ Rev.\ D {\bf 52}, 3081 (1995).

\bibitem{Banks} T.~Banks, M.~Dine and M.~R.~Douglas, hep-ph/0112059.

\bibitem{Gasser}
J.~Gasser and H.~Leutwyler, Phys.\ Rept.\  {\bf 87}, 77 (1982).

\bibitem{Steigman}  For a recent review, see G. Steigman, astro-ph/0009506.

\bibitem{Hogan}  For recent reviews, see
C. J. Hogan, Rev. Mod. Phys. {\bf 72}, 1149 (200);
B. M\" uller, astro-ph/0108259.

\bibitem{Agrawal} 
V.~Agrawal, S.~M.~Barr, J.~F.~Donoghue and D.~Seckel,
Phys.\ Rev.\ Lett.\  {\bf 80}, 1822 (1998)
Phys.\ Rev.\ D {\bf 57}, 5480 (1998).

\bibitem{Livio}  M. Livio, D. Hollowell, A. Weiss and J. W. Truran,
Nature {\bf 340}, 281 (1989).

\bibitem{oberhummer}
A.~Csoto, H.~Oberhummer and H.~Schlattl,
Nucl.\ Phys.\ A {\bf 688}, 560 (2001);
H.~Oberhummer, R.~Pichler and A.~Csoto,
nucl-th/9810057.

\bibitem{Jeltema}  T. Jeltema and M. Sher, Phys. Rev D {\bf 61}, 017301 (2000).

\bibitem{calmet} X.~Calmet and H.~Fritzsch, hep-ph/0112110.

\end{thebibliography}
\end{document}